\documentclass[12pt]{article}
\input epsf.tex
\newcommand{\be}{\begin{equation}}
\newcommand{\ee}{\end{equation}}
\newcommand{\bea}{\begin{eqnarray}}
\newcommand{\eea}{\end{eqnarray}}

\begin{document}

\bigskip 
\begin{titlepage}

\begin{flushright}
UUITP-01/02\\ 
hep-th/0203198
\end{flushright}

\vspace{1cm}

\begin{center}
{\Large\bf A note on inflation and\\
\smallskip

transplanckian physics\\}

\end{center}
\vspace{3mm}

\begin{center}

{\large Ulf H.\ Danielsson

\vspace{5mm}

Institutionen f\"or teoretisk fysik  \\
Box 803, SE-751 08
Uppsala, Sweden}

\vspace{3mm}

{\tt
ulf@teorfys.uu.se\\}

\end{center}

\vspace{5mm}

\begin{center}
{\large \bf Abstract}
\end{center}
\noindent
In this paper we consider the influence of transplanckian physics on the CMBR anisotropies produced by inflation.
We consider a simple toy model that allows for analytic calculations and argue on general grounds,
based on ambiguities in the choice of vacuum,
that effects are expected with a magnitude of the order of $H/\Lambda$, where $H$ is the Hubble constant during
inflation and $\Lambda$ the scale for new physics, e.g. the Planck scale.


\vfill
\begin{flushleft}
March 2002
\end{flushleft}
\end{titlepage}
\newpage


\section{Introduction}

\bigskip

In recent years it has been realized that much can be learnt about the
highest energies and the smallest scales by studying cosmology and in
particular the very early universe. An especially intriguing idea in this
context is inflation. For some nice introductions to inflation with
references see \cite{linde}\cite{liddle}. Inflation successfully solves
several problems of the standard big bang scenario, and also makes a number
of new predictions. Of particular interest is the CMBR anisotropies which
currently is measured with higher and higher precision. Inflation magnifies
tiny quantum fluctuations generated a fraction of a second after the Big
Bang into seeds that eventually cause the formation of galaxies and clusters
of galaxies. The fluctuations leave an imprint on the CMBR that can be used
to test inflation at high precision.

Recently a tantalizing possibility has been discussed in the literature that
suggests that inflation might provide a window towards physics beyond the
Planck scale, [3-24]. Since inflation works by magnifying microscopic
quantum fluctuations into cosmic size, it is reasonable to worry about the
initial linear size of the fluctuations. Were they ever smaller than the
Planck scale? Typically inflation is discussed from a purely field theoretic
perspective, and the only scale in the problem is, basically, the vacuum
energy that generates inflation. As a consequence the quantum fluctuations
are supposed to originate in the infinite past with an infinitely short wave
length. But in the real world we know that fundamentally new physics is to
be expected at the Planck scale, and this simple picture can not be correct.
The key question, then, is whether modifications of the high energy behavior
can change the predictions of inflation. So far no real consensus has been
reached in the literature, and there are at least two competing estimates of
the size of the corrections to the CMBR spectrum. In, e.g., \cite
{Kempf:2001fa} the corrections are argued to be of size $\left( \frac{H}{%
\Lambda }\right) ^{2}$, while in, e.g., \cite{Easther:2001fi}\cite
{Easther:2001fz} one is dealing with substantially larger corrections of
order $\frac{H}{\Lambda }$. $\Lambda $ is the energy scale of new physics,
e.g. the Planck scale or the string scale, and $H$ is the Hubble constant
during inflation. Very recently it was argued in \cite{Kaloper:2002uj},
using a low energy effective field theory, that local physics imply that the
effects can not be larger than $\left( \frac{H}{\Lambda }\right) ^{2}$. This
conclusion has been criticized in \cite{Brandenberger:2002hs}, where it was
pointed out that transplanckian physics can effectively provide the low
energy theory with an excited vacuum, thereby circumventing the arguments of 
\cite{Kaloper:2002uj}.

The purpose of the present paper is to discuss the transplanckian problem
from the point of view of an extremely simple modification of the standard
scenario where we focus on the choice of vacuum. In the usual model of
inflation, the initial state is assumed to be the empty vacuum in the
infinite past when all scales that have a finite linear size today have a
size infinitely smaller than the Planck scale. Even though this does not
make too much sense - after all, we have no idea of how the physics at these
scales work -- it is interesting that this naive approach seems to give
sensible results. To test how robust the predictions are, one has, in the
various works mentioned above, changed by hand the high energy behavior\ in
different ways to see whether and how much the result is influenced. In this
paper we will follow a more conservative approach. We will note when a given
mode reaches a certain minimum scale and encode our ignorance in the choice
of state at that time. For each mode we will choose a time when the mode is
of a specific size comparable to the Planck scale and impose, at that time,
some reasonable initial conditions. A fixed scale for imposing the initial
condition implies that the larger the mode is today, the further back in
time we need to go to impose the initial condition. In fact, one can think
of a semi eternal inflation where modes appear out of nowhere, always at the
same minimum scale, after which inflation takes over and makes them grow
larger. This happens all the time until, for some reason, inflation turns
off. In this way there is a continuous creation of fluctuations with no
unique moment in time (not even the infinite past); it is instead the
unknown small scale physics that produce a certain state at the minimum
scale.

But what is the state created at the minimum scale? The standard proposal
says that it is ``the state that would have been produced if no new physics
occurs on small scales and we start off with an empty vacuum in the infinite
past''. Clearly there is a priori no justification for such a claim. It
might be a reasonable first guess, but it should be the subject of criticism
and discussion. Another proposal, no less plausible or implausible, would be
that the state produced at the minimum scale is the vacuum determined by
some principle of naturalness at \textit{that} time. The idea would then be
that the span provided by the two choices gives a reasonable estimate of how
uncertain our prediction of the fluctuation spectrum is due to
transplanckian physics. That is, how sensitive inflation is to reasonable
variations in the initial conditions.

In agreement with \cite{Brandenberger:2002hs} we will find corrections
larger than those discussed in \cite{Kaloper:2002uj}. Similar ideas as those
presented in this paper can also be found in \cite{Easther:2001fi}\cite
{Easther:2001fz}, but our analysis will be made in a simpler model that
allows for analytic results and provides some further insight into what the
nature of the effect is. In particular, we will discuss the role of the
adiabatic vacuum. In fact, our main result will be that \textit{the} \textit{%
expected magnitude of the transplanckian corrections will be given by the
magnitude of the first order corrections to the zeroth order adiabatic vacuum%
}.

\bigskip

\section{A simple model}

\bigskip

\subsection{A Heisenberg setup}

\bigskip

Let us consider an inflating background with metric 
\begin{equation}
ds^{2}=dt^{2}-a\left( t\right) ^{2}dx^{2},
\end{equation}
where the scalefactor is given by $\ a\left( t\right) =e^{Ht}.$ The equation
for a scalar field in this background is given by 
\begin{equation}
\ddot{\phi}+3H\dot{\phi}-\nabla ^{2}\phi =0.
\end{equation}
In terms of the conformal time $\eta =-\frac{1}{aH}$, and the rescaled field 
$\mu =a\phi $, we find 
\begin{equation}
\mu _{k}^{\prime \prime }+\left( k^{2}-\frac{a^{\prime \prime }}{a}\right)
\mu _{k}=0  \label{modeeq}
\end{equation}
in Fourier space. Prime refers to derivatives with respect to conformal
time. Note that we have $k=ap$, where $p$ is the physical momentum which is
redshifting away with the expansion ($k$ is fixed). We will also need the
conjugate momentum to $\mu _{k}$ which is given by: 
\begin{equation}
\pi _{k}=\mu ^{\prime }-\frac{a^{\prime }}{a}\mu _{k}.
\end{equation}
When quantizing the system it turns out that the Heisenberg picture is the
most convenient one to use. A nice discussion of this approach can be found
in \cite{Polarski:1995jg}, see also \cite{liddle}. In terms of time
dependent oscillators we can write 
\begin{eqnarray}
\mu _{k}\left( \eta \right)  &=&\frac{1}{\sqrt{2k}}\left( a_{k}\left( \eta
\right) +a_{-k}^{\dagger }\left( \eta \right) \right)   \nonumber \\
\pi _{k}\left( \eta \right)  &=&-i\sqrt{\frac{k}{2}}\left( a_{k}\left( \eta
\right) -a_{-k}^{\dagger }\left( \eta \right) \right) .
\end{eqnarray}
The oscillators can be conveniently expressed in terms of their values at
some fixed time $\eta _{0}$, 
\begin{eqnarray}
a_{k}\left( \eta \right)  &=&u_{k}\left( \eta \right) a_{k}\left( \eta
_{0}\right) +v_{k}\left( \eta \right) a_{-k}^{\dagger }\left( \eta
_{0}\right)   \label{oscutv} \\
a_{-k}^{\dagger }\left( \eta \right)  &=&u_{k}^{\ast }\left( \eta \right)
a_{-k}^{\dagger }\left( \eta _{0}\right) +v_{k}^{\ast }\left( \eta \right)
a_{k}\left( \eta _{0}\right) ,  \nonumber
\end{eqnarray}
which is nothing but the Bogolubov transformations which describes the
mixing of the creation and annihilation operators as time goes by. Plugging
this back into the expressions for $\mu _{k}\left( \eta \right) $ and $\pi
_{k}\left( \eta \right) $ we find: 
\begin{eqnarray}
\mu _{k}\left( \eta \right)  &=&f_{k}\left( \eta \right) a_{k}\left( \eta
_{0}\right) +f_{k}^{\ast }\left( \eta \right) a_{-k}^{\dagger }\left( \eta
_{0}\right)   \nonumber \\
\pi _{k}\left( \eta \right)  &=&-i\left( g_{k}\left( \eta \right)
a_{k}\left( \eta _{0}\right) -g_{k}^{\ast }\left( \eta \right)
a_{-k}^{\dagger }\left( \eta _{0}\right) \right) ,
\end{eqnarray}
where 
\begin{eqnarray}
f_{k}\left( \eta \right)  &=&\frac{1}{\sqrt{2k}}\left( u_{k}\left( \eta
\right) +v_{k}^{\ast }\left( \eta \right) \right)   \nonumber \\
g_{k}\left( \eta \right)  &=&\sqrt{\frac{k}{2}}\left( u_{k}\left( \eta
\right) -v_{k}^{\ast }\left( \eta \right) \right) .
\end{eqnarray}
$f_{k}\left( \eta \right) $ is a solution of the mode equation (\ref{modeeq}%
). We are now in the position to start discussing the choice of vacuum. A
reasonable candidate for a vacuum is

\begin{equation}
a_{k}\left( \eta _{0}\right) \left| 0,\eta _{0}\right\rangle =0.
\label{defvac}
\end{equation}
In general this corresponds to a class of different vacua depending on the
parameter $\eta _{0}$. \ At this initial time it follows from (\ref{oscutv})
that $v_{k}\left( \eta _{0}\right) =0$, and the relation between the field
and its conjugate momentum is particularly simple: 
\begin{equation}
\pi _{k}\left( \eta _{0}\right) =ik\mu _{k}\left( \eta _{0}\right) .
\end{equation}
This choice of vacuum has a simple physical interpretation. Following \cite
{Polarski:1995jg} it is easy to show that it corresponds to a state which
minimizes the uncertainty at $\eta =\eta _{0}$. \ Using $\left\langle \mu
_{k}\right\rangle =\left\langle \pi _{k}\right\rangle =0$ it follows that 
\begin{eqnarray}
\left\langle \Delta \mu _{k}\Delta \mu _{k^{\prime }}\right\rangle 
&=&\left| f_{k}\right| ^{2}\delta ^{(3)}\left( \mathbf{k}-\mathbf{k}^{\prime
}\right)   \nonumber \\
\left\langle \Delta \pi _{k}\Delta \pi _{k^{\prime }}\right\rangle 
&=&\left| g_{k}\right| ^{2}\delta ^{(3)}\left( \mathbf{k}-\mathbf{k}^{\prime
}\right) ,
\end{eqnarray}
where 
\begin{equation}
\left| f_{k}\right| ^{2}\left| g_{k}\right| ^{2}=\frac{1}{4}\left( 1+\left|
uv-u^{\ast }v^{\ast }\right| ^{2}\right) .
\end{equation}
The latter expression is indeed minimized at $\eta =\eta _{0}$ where $%
v_{k}\left( \eta _{0}\right) =0$.

We will now show that the vacuum defined in this way can be referred to as
the zeroth order adiabatic vacuum.

\bigskip

\subsection{The role of the adiabatic vacuum}

In a time dependent background the notion of a vacuum is a tricky issue. The
ideal situation is if the there is only some transitional time dependence,
in which case there is a unique definition of the vacuum in the infinite
past as well as in the infinite future. The time evolution of the initial
vacuum will, however, not necessarily generate the final vacuum, a
phenomenon which we interpret as the creation of particles. With a time
dependence that never shuts off, the situation is less clear. One
possibility is to use the adiabatic vacuum, where the solution of the wave
equation is, formally, assumed to be of WKB-form. A nice discussion of the
adiabatic vacuum can be found in \cite{birrell}. Often the exact solution is
expanded to some finite order in the adiabatic parameter that determines the
slowness of the process. The idea is to approximate the field equations, at
some moment in time, with their time independent counterparts (possibly with
corrections to some finite order) and define positive and negative energy
using solutions to these approximative equations. Even though the adiabatic
vacuum obtained in this way in general does not correspond to a solution of
the exact field equation, it certainly corresponds to \textit{some }specific
choice of vacuum. What one should remember, however, is that the adiabatic
vacuum (to some finite order in the adiabatic parameter) is not unique but
depends on what moment in time one uses for its definition. In de Sitter
space, however, it happens that the finite order adiabatic vacuum obtained
in the infinite past actually corresponds to an exact solution of the exact
field equations, and therefore in some sense is distinguished. After all,
when the modes are small enough they do not care about the expansion of the
universe.

Which vacuum should we choose? One possibility is to use the adiabatic
vacuum of arbitrary order -- corresponding to an exact solution -- but there
are also other choices like the one of minimum uncertainty discussed in the
previous section. As we will argue below the minimum uncertainty vacuum
agrees with the adiabatic one only to zeroth order. In fact, it is only at
zeroth order, where the expansion of the universe can be ignored, that
ambiguities in the definition of the vacuum are removed. It is important to
observe that these distinctions between various vacua only become important
since we insist on imposing the choice of vacua at a finite time
corresponding to some specific finite wavelength, e.g. the Planck scale. Any
claim about the structure of the vacuum beyond the zeroth order, needs
knowledge of physics on this scale. Since such knowledge is currently not
available, we can only list various alternatives. The vacuum choice of the
previous section represents one such viable alternative, besides the
standard one, and can be used to indicate the natural span of possibilities.
Let us now proceed with a more detailed analysis.

In the zeroth order adiabatic approximation, the solutions of a mode
equation of the form 
\begin{equation}
\mu _{k}^{\prime \prime }+\left( k^{2}-C\left( \eta \right) \right) \mu
_{k}=0,
\end{equation}
is given by 
\begin{equation}
\mu _{k}=\frac{1}{\sqrt{2\omega }}e^{\pm i\omega \eta },
\end{equation}
where 
\begin{equation}
\omega =\sqrt{k^{2}-C\left( \eta \right) }.
\end{equation}
For the approximation to make sense we must have an $\omega $ that varies
slow enough (i.e. adiabatically). A necessary condition for this to be the
case is that 
\begin{equation}
\frac{d}{d\eta }\ln C\ll \omega ,
\end{equation}
which for us (where $C\left( \eta \right) =\frac{2}{\eta ^{2}}$) typically
leads to 
\begin{equation}
k\eta \gg 1.
\end{equation}
With the help of this the zeroth order solution simply degenerates into 
\begin{equation}
\mu _{k}=\frac{1}{\sqrt{2k}}e^{\pm ik\eta },
\end{equation}
and one finds a conjugate momentum given by 
\begin{equation}
\pi _{k}=ik\mu _{k}.  \label{pimu}
\end{equation}
This is precisely what our choice in the previous subsection led to, and we
can therefore refer to the vacuum that we will analyze as the zeroth order
adiabatic vacuum. As pointed out above, a finite order adiabatic mode is in
general not an exact solution of the field equations, but the vacuum that it
corresponds to is nevertheless an honest proposal for a vacuum. One should
view (\ref{pimu}) as initial conditions with a subsequent time evolution
given by the exact solution.

\subsection{Imposing the initial conditions}

\bigskip

Let us now consider the standard treatment of fluctuations in inflation. In
this case we have 
\begin{equation}
f_{k}=\frac{1}{\sqrt{2k}}e^{-ik\eta }\left( 1-\frac{i}{k\eta }\right) 
\end{equation}
and 
\begin{equation}
g_{k}=\sqrt{\frac{k}{2}}e^{-ik\eta }.
\end{equation}
The logic behind the choice is that the mode at early times (when $\eta
\rightarrow -\infty $) is of positive frequency and corresponds to what one
would naturally think of as the vacuum. It is nothing but the state obeying (%
\ref{defvac}) for $\eta _{0}\rightarrow -\infty $ and is therefore the
zeroth order adiabatic vacuum of the infinite past. Note that the zeroth
order adiabatic vacuum in this case is actually an exact solution (for $\eta
\rightarrow -\infty $). For later times (when $\eta \rightarrow 0$ and the
second term of $f_{k}$ becomes important) we see how the initial vacuum
leads to particle creation thereby providing the fluctuation spectrum.

But what if the initial conditions are chosen differently? In general we
could have 
\begin{eqnarray}
f_{k} &=&\frac{A_{k}}{\sqrt{2k}}e^{-ik\eta }\left( 1-\frac{i}{k\eta }\right)
+\frac{B_{k}}{\sqrt{2k}}e^{ik\eta }\left( 1+\frac{i}{k\eta }\right)  
\nonumber \\
g_{k} &=&A_{k}\sqrt{\frac{k}{2}}e^{-ik\eta }-B_{k}\sqrt{\frac{k}{2}}%
e^{ik\eta },
\end{eqnarray}
with a nonzero $B_{k}$. If we then work backwards, we can calculate what
this corresponds to in terms of $u_{k}$ and $v_{k}$. The result is: 
\begin{eqnarray}
u_{k} &=&\frac{1}{2}\left( A_{k}e^{-ik\eta }\left( 2-\frac{i}{k\eta }\right)
+B_{k}e^{ik\eta }\frac{i}{k\eta }\right)   \nonumber \\
v_{k}^{\ast } &=&\frac{1}{2}\left( B_{k}e^{ik\eta }\left( 2+\frac{i}{k\eta }%
\right) -A_{k}e^{-ik\eta }\frac{i}{k\eta }\right) .
\end{eqnarray}
At this point we should also remember that 
\begin{equation}
\left| u_{k}\right| ^{2}-\left| v_{k}\right| ^{2}=1
\end{equation}
from which we find 
\begin{equation}
\left| A_{k}\right| ^{2}-\left| B_{k}\right| ^{2}=1.
\end{equation}
As we have seen, the choice of vacuum that we make requires that we put $%
v_{k}^{\ast }\left( \eta _{0}\right) =0$ at some initial moment $\eta _{0}$.
This implies that 
\begin{equation}
B_{k}=\frac{ie^{-2ik\eta _{0}}}{2k\eta _{0}+i}A_{k},
\end{equation}
from which we conclude that 
\begin{equation}
\left| A_{k}\right| ^{2}=\frac{1}{1-\left| \alpha _{k}\right| ^{2}},
\end{equation}
where 
\begin{equation}
\alpha _{k}=\frac{i}{2k\eta _{0}+i}.
\end{equation}

\bigskip

We next move to the calculation of the fluctuation spectrum given by: 
\begin{eqnarray}
P_{\phi } &=&\frac{1}{a^{2}}P_{\mu }=\frac{k^{3}}{2\pi ^{2}a^{2}}\left|
f_{k}\right| ^{2}\sim \frac{1}{4\pi ^{2}\eta ^{2}a^{2}}\left( \left|
A_{k}\right| ^{2}+\left| B_{k}\right| ^{2}-A_{k}^{\ast
}B_{k}-A_{k}B_{k}^{\ast }\right)   \nonumber \\
&=&\left( \frac{H}{2\pi }\right) ^{2}\left( 1+\left| \alpha _{k}\right|
^{2}-\alpha _{k}e^{-2ik\eta _{0}}-\alpha _{k}^{\ast }e^{2ik\eta _{0}}\right) 
\frac{1}{1-\left| \alpha _{k}\right| ^{2}},
\end{eqnarray}
where we have used $\eta =-\frac{1}{aH}$ in the prefactor and considered the
leading term at late times when $\eta \rightarrow 0$. If we impose the
initial condition at $\eta _{0}\rightarrow -\infty $ we get $\alpha =0$ and
recover the standard result $P_{\phi }=\left( \frac{H}{2\pi }\right) ^{2}$.
But let us now do something different following the discussion in the
introduction. For a given $k$ we choose a finite $\eta _{0}$ such that the
physical momentum corresponding to $k$ is given by some fixed scale $\Lambda 
$. $\Lambda $ is the energy scale of new physics, e.g. the Planck scale or
the string scale. From 
\begin{equation}
k=ap=-\frac{p}{\eta H}
\end{equation}
with $p=\Lambda $ we find 
\begin{equation}
\eta _{0}=-\frac{\Lambda }{Hk}.
\end{equation}
It is important to note that $\eta _{0}$ depends on $k$. If we assume $\frac{%
\Lambda }{H}\gg 1$ we get 
\begin{equation}
P_{\phi }=\left( \frac{H}{2\pi }\right) ^{2}\left( 1-\frac{H}{\Lambda }\sin
\left( \frac{2\Lambda }{H}\right) \right) ,
\end{equation}
which is our final result.\footnote{%
If the field that we are considering is a gravitational mode, $P_{\phi }$
directly gives the density fluctuations. For a scalar field, on the other
hand, one needs to take an extra factor $\left( \frac{H}{\dot{\phi}}\right)
^{2}$ into account. See \cite{liddle} for further details.}

\subsection{Some comments on the result}

There are several comments one can make. First, one verifies that the size
of the correction ($\sim \frac{H}{\Lambda }=\left| \frac{1}{k\eta _{0}}%
\right| $) is precisely what to be expected from a higher order correction
to the zeroth order adiabatic vacuum. If the vacuum is imposed in the
infinite past, the vacuum is exact, but if it is imposed at a later time it
is natural to expect nonvanishing corrections. Corrections of precisely this
order of magnitude have been found in, e.g., \cite{Easther:2001fi}\cite
{Easther:2001fz} and as we have seen this can be expected on quite general
grounds. These corrections are in general larger than those discussed in 
\cite{Kaloper:2002uj} which went like $\left( \frac{H}{\Lambda }\right)
^{2}. $

Second, it is interesting to note that if we have a model of inflation where 
$H$ is slowly changing, leading to a spectrum which is not exactly scale
invariant, the correction term will be very sensitive to $k$ through the
dependence of $H$ on $k$. That is, there will be a modulation of $P_{\phi }$%
. In fact, one could expect this to be a rather general phenomena in models
where the initial conditions are set at a particular scale. The modulation
that we have found is precisely of the same form as in the numerical work of 
\cite{Easther:2001fz} which considered a specific example of slow roll. It
would be interesting to study this phenomenon in a more systematic way for
various models.

\section{Summary and conclusions}

In this paper we have studied the possible influence of transplanckian
physics on the fluctuation spectrum predicted by inflation. We have made use
of an extremely natural initial condition: we require that the modes are
created in a state of minimized uncertainty. If this is imposed in the
infinite past there is no difference between this choice and the usual
choice of an adiabatic vacuum. But contrary to the standard treatment we
have imposed the initial condition not in the infinite past, but at a mode
dependent time determined by when a particular mode reaches a size of the
order of the fundamental scale (e.g. the Planck scale). As a consequence our
analysis agrees with the standard choice only to zeroth order in an
adiabatic expansion with corrections at first order. This should be viewed
as a conservative approach appropriate for estimating how well the
fluctuation spectrum can be predicted without any knowledge of high energy
physics. To phrase it differently: if measurements can be done at the
accuracy required, transplanckian physics will be within reach. The size of
the prediction is not large\footnote{%
If $\Lambda $ is the Planck scale, $\frac{H}{\Lambda }$ is at most $10^{-4}$%
. If $\Lambda $ is the string scale, $\frac{H}{\Lambda }$ could possibly be $%
10^{-2}$ in a very optimistic scenario.} but it would be interesting to
further analyze under what circumstances it might be observable.

In this context one should also consider the bound found in \cite
{Starobinsky:2001kn}. There it is argued that present day physics severely
limits how far from the standard vacuum choice one can deviate. The problem
is that with a large deviation too many particles (e.g. gravitons) will be
produced which could contribute to the present energy density. According to
this estimate the coefficient in front of the wrong mode can be at most of
the order $\frac{H_{0}}{\Lambda }$, where $H_{0}$ is the Hubble constant
now. But, as argued in \cite{Easther:2001fi}, when the coefficient is traced
back in time it might very well correspond to a considerably larger ratio in
the past. In fact, a natural expectation is that it becomes of the order $%
\frac{H}{\Lambda }$, meaning that it is really a tricky question involving
numbers of order not too far from one. Similar comments applies to the work
of \cite{Tanaka:2000jw}, which discusses back reaction due to particle
production during inflation itself. More detailed discussions on these
issues can be found in \cite{Brandenberger:2002hs}. Actually, one way to
view the argument of \cite{Starobinsky:2001kn} is as yet another example of
how sensitive inflation is to transplanckian physics.

The conclusion is, therefore, that effects of transplanckian physics are
possibly within the reach of cosmological observations even though much more
detailed calculations are required to make a definite statement. But even
this is much more optimistic than the usual expectations in standard
particle physics, and could imply a very exciting future for cosmology as
well as string theory.

\bigskip

\section*{Acknowledgments}

I would like to thank Robert Brandenberger for valuable discussions. The
author is a Royal Swedish Academy of Sciences Research Fellow supported by a
grant from the Knut and Alice Wallenberg Foundation. The work was also
supported by the Swedish Research Council (VR).

\end{document}